# Core Placement Optimization of Many-core Brain-Inspired Near-Storage Systems for Spiking Neural Network Training

**Xueke Zhu[1], Wenjie Lin[1†], Yanyu Lin[1], Yunhao Ma[1,3], Wenxiang Cheng[1], Zhengyu Ma[1], Yonghong Tian[1,2], Huihui Zhou[1†]**

**[1]Pengcheng Laboratory, [2]Peking University,**

**[3]Southern University of Science and Technology**

**{zhuxk, linwj, zhouhh}@pcl.ac.cn**

**Abstract**: With the increasing application scope of spiking neural networks (SNN), the complexity of SNN models has surged, leading to an exponential growth in demand for AI computility. As the new generation computing architecture of the neural networks, the efficiency and power consumption of distributed storage and parallel computing in the many-core near-memory computing system have attracted much attention. Among them, the mapping problem from logical cores to physical cores is one of the research hotspots. In order to improve the computing parallelism and system throughput of the many-core near-memory computing system, and to reduce power consumption, we propose a SNN training many-core deployment optimization method based on Off-policy Deterministic Actor-Critic. We utilize deep reinforcement learning as a nonlinear optimizer, treating the many-core topology as network graph features and using graph convolution to input the many-core structure into the policy network. We update the parameters of the policy network through near-end policy optimization to achieve deployment optimization of SNN models in the many-core near-memory computing architecture to reduce chip power consumption. To handle large-dimensional action spaces, we use continuous values matching the number of cores as the output of the policy network and then discretize them again to obtain new deployment schemes. Furthermore, to further balance inter-core computation latency and improve system throughput, we propose a model partitioning method with a balanced storage and computation strategy. Our method overcomes the problems such as uneven computation and storage loads between cores, and the formation of local communication hotspots, significantly reducing model training time, communication costs, and average flow load between cores in the many-core near-memory computing architecture.

## 1 INTRODUCTION

In recent years, with the significant development of deep learning programming frameworks[1,2], deep neural networks have been more widely applied in complex tasks such as computer vision[3,4], speech recognition[5], natural language processing[6], automatic driving[4], and recommendation systems[7]. Spiking neural networks (SNN) as a new generation neural network, have attracted considerable attention due to their high biological plausibility, event-driven nature, and low power consumption[8]. Deep spiking neural network models adopt discrete binary activation and spatiotemporal backpropagation, making the demand for computing resources more prominent during the training process. Therefore, the hardware architecture for SNN model training requires higher resource efficiency.

†Corresponding author

Some SNN learning chips use local synaptic plasticity such as STDP for weight updates. The absence of backpropagation facilitates their implementation on distributed many-core neural morphology architectures. Although they offer low power consumption and fast response, they still suffer from low precision issues[9-11]. On the other hand, architectures like Loihi[12] and H2Learn[13] are designed to ensure high precision in SNNs. For example, H2Learn is based on the sparsity of spiking neural networks and is equipped with forward, backward, and weight update engines, enabling efficient training of SNN models based on BPTT. As tasks become increasingly complex and networks grow deeper with larger computations, many-core architectures have attracted attention. For instance, the Tianjic[14,15] is a many-core architecture that supports both spiking neural networks and convolutional recurrent neural networks. Each core can be configured as a neural network unit, utilizing parallel Multiply-Accumulate (MAC) units for efficient and flexible computations to perform deep SNN model inference. However, the Tianjic does not support SNN model training, which still heavily relies on GPUs. Although GPUs excel in parallel computing, they are not suitable for training SNN models due to factors such as the event-driven computing mechanism, low power consumption, high performance, and von Neumann bottleneck associated with spiking neural networks. Table 1 shows the relevant test data for inference of two SNN models on GPU and HP300 (Commercial version of Tianjic). The data indicates that GPUs exhibit very low EER for SNN model computations. Therefore, exploring many-core near-memory computing architectures is crucial for achieving training of deep spiking neural network models.

Table 1: The SNN model inference performance testing

| Spike Models | Batch Size | Device Name | Frame Rate(fps) | Power(W) | EER(fps/W) |
|---|---|---|---|---|---|
| Spike-Unet | 1 | GPU V100 | 43.70 | 106.52 | 0.41 |
| | | Lynxi HP300 | 126.00 | 16.85 | 7.48 |
| Spike-ResNet50 | 1 | GPU V100 | 17.50 | 72.19 | 0.24 |
| | | Lynxi HP300 | 37.00 | 15.50 | 2.38 |

The many-core near-memory computing system is a novel computing system architecture that offers significant advantages in distributed storage and parallel computing. However, it faces substantial challenges in physical mapping, model compilation, power consumption, and other aspects. Among these challenges, the mapping problem from logical cores to physical cores is one of the hot research topics, as depicted in Figure 1. Myun[16] has proposed an optimization method for ANN model inference deployment in multi-chip many-core systems based on reinforcement learning. This method leverages the advantages of reinforcement learning in solving combinatorial optimization problems to select the optimal mapping scheme from logical cores to physical cores, thereby reducing model inference latency and lowering power consumption.

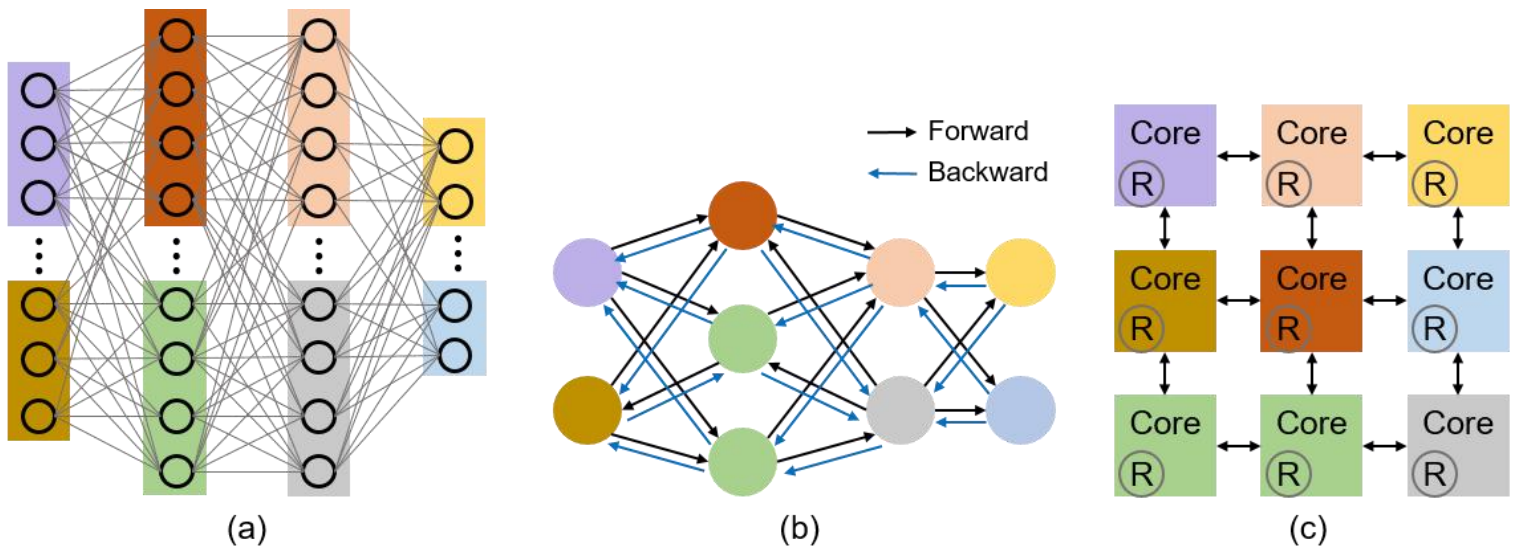

Figure 1. Model-to-core mapping process: (a)neural network; (b) neural network is partitioned and is represented as a weighted DAG; (c) logic cores are one-to-one mapped into physical nodes.

Research on neural network model training accelerators mainly focuses on scheduling and allocation between data flow and hardware resources. Data flow includes model partitioning strategies, computation orders, and pipelining methods, while hardware resource scheduling mainly involves computing resources and memory. Most studies concentrate on designing efficient data flow strategies to enhance data reuse. For the same model, various partitioning strategies and resource allocation schemes can be applied. For example, NVDLA and Eyeriss[17] adopt different data flow strategies, while frameworks like MAESTRO[18] and Timeloop[19] select the optimal data flow by calculating data movement and computational operation power consumption. Given the data flow and hardware resources, different partitioning strategies and resource allocations may result in significant differences in latency and energy consumption during neural network inference or training[20]. Currently, most optimization methods for hardware resource allocation are based on exhaustive search[21-23], which are inefficient. Confuciux[24] utilizes reinforcement learning techniques to adaptively allocate hardware resources for deep neural network accelerators, thereby improving the acceleration effect of deep neural networks through automatic resource allocation.

This paper mainly investigates the logical partitioning of spiking neural network cores and the physical mapping method for training in the many-core near-memory computing architecture. Addressing issues such as uneven inter-core computation and storage loads, susceptibility to local communication hotspots, low utilization of computing cores, and long model training times in the training process of deep spiking neural networks, we utilize a deep reinforcement learning network as a nonlinear optimizer. Through deep deterministic policy gradient, we automatically optimize the deployment network of many-core systems, proposing an optimization method from the logical partitioning of SNN model cores to the mapping of models in many-core near-memory computing architectures. The main contributions of this paper are as follows:

1) Taking into account the storage and computing resources of the many-core near-memory computing architecture, we propose a logical core partitioning method for deep SNN models, aiming to balance the computation latency of each logical core block on the target physical core. This is to avoid the "bucket effect" caused by balanced computation or storage partitioning and significantly improve the efficiency of the computing system. Experimental results demonstrate the effectiveness of our approach in enhancing performance.

2) Mapping the logical-to-physical core mapping optimization in SNN model training as a reinforcement learning problem, we transform the connectivity between logical cores into a logical graph, with the topology of the many-core architecture as the network graph. Utilizing graph convolution, we embed the features of the logical graph into the policy network. The policy network outputs continuous values matching the number of logical cores, which are discretized to obtain deployment schemes with low data transfer volume and short computation latency.

3) In terms of parallel pipelining computation, we introduced a fine-grained computation pipelining method based on FPDeep[25], which specializes the convolution operation for the memory hierarchy and inter-core communication structure of the deep pipeline computation. By parallelizing the computation between different convolutional layers, we aim to enhance the computational parallelism of model, improve chip utilization, and reduce training time.

## 2 RELATED WORK

### 2.1 Neuromorphic Accelerator

Efficient training of spiking neural networks (SNNs) has been a topic of great interest in recent years. GPUs remain the primary platform for training SNNs, despite their hardware architecture and programming operator libraries not fully leveraging the advantages of SNN data formats and computational paradigms. In recent years, researchers have developed specific accelerators for SNN training. FlexLearn[26] is an on-chip learning engine that realizes an end-to-end simulation accelerator for SNNs. Loihi[12] is a neuromorphic many-core processor with adaptive learning capabilities, enabling efficient training of SNNs directly on the chip. H2Learn[13] is an efficient SNN learning accelerator architecture based on Backpropagation Through Time (BPTT). This architecture enhances overall performance by designing forward engines, backward engines, and weight update engines.

### 2.2 Model Segmentation

In many-core computing architecture, model segmentation is a strategy designed to make full use of the characteristics of many-core processors. By allocating different parts of a deep learning model to different processing cores for execution, the training or inference speed of the model can be significantly improved, so different model segmentation methods are constantly emerging. The horizontal segmentation method of the model[27,28] is used to allocate different sample data of the model to many-core devices for parallel computing. In horizontal segmentation, data and model parameters need to be frequently exchanged between different devices, and a large amount of data transmission becomes a performance bottleneck and affects the training efficiency. These two algorithms[29,30] use vertical segmentation of the model (splitting by layer) to split the model weight of the physical mapping, and allocate different parts of the model to different devices or nodes for parallel computing according to the structure, function and computing requirements of the model to achieve higher computational efficiency and performance. However, due to the different amount of computation and intermediate data in different layers of the deep learning model, the amount of computation and intermediate data in different layers of the deep learning model are different. Vertical partitioning is prone to the problem of load imbalance between cores, which reduces the overall system performance. To further optimize partitioning methods, Core Placement[7] evenly partitions the model along input and output channels to balance the computational tasks of each sub-model.

### 2.3 Mapping Optimization

Mapping deep SNN models involves two processes: logical mapping and physical mapping. Currently, physical mapping is garnering significant attention in research. The Prime architecture[31] which utilizes ReRAM as dual-function memory and computing units, placing computational functions in the main memory to reduce data transfer requirements and greatly improve computational efficiency and energy efficiency. SemiMap[32] is a framework for achieving a balance between speed and overhead in convolutional operations. This method folds physical resources along the row dimension of feature maps (FM) while expanding them along the column dimension, reducing resource usage and increasing parallelism. However, these studies did not address optimization issues in physical mapping. In the design of TrueNorth[33], various global layout algorithms are employed, including multi-level partition-driven algorithms, analytic constraint generation algorithms, hierarchical quadratic placement algorithms, and force-directed placement algorithms based on quartic curves. These algorithms optimize the layout of the circuit by mapping logically connected cores to physically adjacent cores, thereby reducing wire lengths within and between chips and lowering power consumption. IBM CPLACE[34] proposed an algorithm called RQL for global layout, which obtains high-quality circuit layout solutions through quadratic wire length optimization and fixed-point-based module extension. The RQL algorithm gradually reduces module overlap iteratively to achieve better layout results. Wu[7] introduced a reinforcement learning-based deployment scheme to improve system performance during model inference on multi-chip many-core architectures. The above reinforcement learning-based deployment optimization schemes are only applied to neural network inference and have not explored the training aspect of the model.

System deployment optimization in many-core systems is essentially a combinatorial optimization problem, aiming to find the optimal or near-optimal solutions under finite resources, it typically exhibits high complexity and challenges. Reinforcement learning algorithms demonstrate significant potential in solving such problems by learning optimal strategies through interaction with the environment[35]. In the design of heterogeneous many-core architectures, layout optimization is a complex and time-consuming process, constrained by various factors including the size of input feature maps, inter-core topology, and properties of computational cores. A deployment optimization method[36] based on a combination of cyclic reinforcement learning and simulated annealing algorithms, which integrates the exploratory nature of reinforcement learning with the local search capabilities of simulated annealing, to iteratively optimize and obtain the best deployment solutions. Experiments have shown that this method effectively improves the performance and efficiency of integrated circuits. The DeepChip[37] transforms the problem of chip multicore deployment into a reinforcement learning problem, learning interactively with the environment to generate high-quality deployment solutions, experiment results demonstrate the significant improvements in the performance and efficiency of multicore chips. Additionally, an adaptive optimization method[38] for deploying neural network chips based on policy gradient reinforcement learning, utilizing recurrent neural networks as the policy networks to predict placements and iteratively optimize deployment solutions. Spotlight[39] presented a customized near-end policy optimization algorithm, aimed at addressing the deployment problem in training deep neural networks. Experimental results show that Spotlight utilizes the partial connectivity of convolutional neural networks to balance the computation load on GPUs without incurring any communication overhead. Furthermore, a device placement algorithm based on near-end policy optimization, which selects the optimal deployment strategy by minimizing cross-entropy[40]. Baechi[41] was a rapid deployment method, which dynamically allocating operations among different cores to maximize parallel computation and reduce communication overhead.

## 3 PROBLEM STATEMENT

### 3.1 Physical Mapping

Physical mapping is a process that aims to accurately map logical graphs onto the physical cores of a many-core system. These logical graphs, essentially task diagrams for applications, are constructed based on the connectivity directions and communication weights among model chunks. These logical graphs will be deployed and scheduled on the physical cores to achieve efficient parallel processing.

As in the Policy[16], we have developed a series of mathematical definitions for these modeling elements, enabling a more precise description and analysis of the physical mapping process.

**Definition A**: An application task graph is defined as $DAG$, denoted as $M(A,E)$, where $a_i \in A$ represents the set of tasks, $a_i$ represents one of the subtasks, and$e_{i,j} \in E$ represents the connection direction communication weights between tasks $a_i$ and $a_j$.

**Definition B**: The NOC topology graph is a directed graph $T(N,P)$, where each node in the directed graph $T$ is connected to a computing node in the many-core system, and each path $p_{i,j} \in P$ represents the shortest data transmission path from node $n_i$ to node $n_j$ under the routing policy.

**Definition C**: A mathematical model that maps the application task graph $M(A,E)$ to the NoC topology grap $T(N,P)$is defined by the following function:

$$Map: M \to T \ \boldsymbol{s.t.}\ map(a_i) = n_i, \forall a_i \in A, \exists n_i \in N, |A| \leq |N| \tag{1}$$

Where $|A|$ and $|N|$ represent the number of application tasks and routers of physical device.

### 3.2 Routing Policy

Routing strategies play a pivotal role in computer networks, profoundly influencing the transmission trajectory of data packets. Depending on various needs and application scenarios, we can flexibly choose from a diverse range of routing strategies, each with its unique advantages and applicable areas. For instance, the shortest path routing strategy aims to find the most direct and efficient route for data packets to reach their destination, while the least hop routing strategy focuses on selecting paths with the minimum number of intermediate nodes to reduce transmission delays. In addition, there are more sophisticated routing strategies, such as dynamic routing and static routing. Dynamic routing strategies can perceive real-time changes in network traffic and adjust routing paths accordingly to ensure efficient and stable network operation. On the other hand, static routing strategies rely on predefined rules for path selection, suitable for scenarios where the network structure is relatively stable and traffic changes are minimal. In this paper, we have adopted a clockwise search-based shortest path routing strategy. This strategy not only effectively ensures network stability but also significantly enhances data transmission efficiency, providing strong support for building efficient and stable computer networks.

### 3.3 Reinforcement Learning

Reinforcement learning is a method that enables agents to learn optimal action strategies through interaction with their environment. Unlike traditional supervised and unsupervised learning, reinforcement learning emphasizes learning based on cumulative feedback from the environment, with agents continuously adjusting their strategies based on their actions and the state of the environment. During Reinforcement Learning Model working, an agent generates an action$a_t$based on the state $s_t$ received from the environment, executes that action, and interacts with the environment. The environment then provides feedback $r_t$ (i.e., a reward signal) based on the agent's performance, evaluating the quality of the executed action. Simultaneously, the environment transitions to the next state $s_{t+1}$. The ultimate goal is to learn an optimal strategy that enables the agent to maximize cumulative rewards in a given environment, thereby making optimal decisions.

The value function is used to evaluate how good or bad it is to take an action in a given state. The value function is defined as Equation (2):

$$V_Func_{(s)} = E_{\pi}[R_t | S_t = s] \tag{2}$$

The state-action q-value function $Q(s, a)$ of action $a$ in state $s$ is expressed as follows Equation (3):

$$Q_{\pi}(s, a) = E_{\pi}[R_t | S_t = s, A_t = a] \tag{3}$$

We employ the Proximal Policy Optimization (PPO) algorithm to update our strategies. Firstly, initializing a baseline strategy and repeatedly execute it in the environment to explore. During the exploration process, we collect data to evaluate the value of the strategy and then utilize the gradient descent method to optimize the strategy, aiming to enhance its value. Notably, the PPO algorithm incorporates a technique known as "clip" which effectively prevents the strategy from becoming overly aggressive by limiting the magnitude of its updates. We continuously repeat these steps until the strategy reaches its optimal state. The approach performs excellently in the process of strategy optimization, providing robust support for improving system performance and stability.

# 4 METHODOLOGY

This chapter initially introduces the background information necessary for modeling the optimization system for deploying the neuromorphic computing system with multiple cores and proximity memory, including physical mapping, routing strategies, and reinforcement learning models. Finally, we elaborate on the optimization method for deploying multiple cores based on Actor-Critic pulse neural network training.

## 4.1 The Architecture of Many-Core Near-Memory Brain-Inspired Systems

We design a many-core near-memory brain-like computing architecture that supports FP, BP, and WG computations for deep SNN training, as shown in Figure 2. The single computing core within the many-core architecture includes FP Engine, BP Engine, and Router.

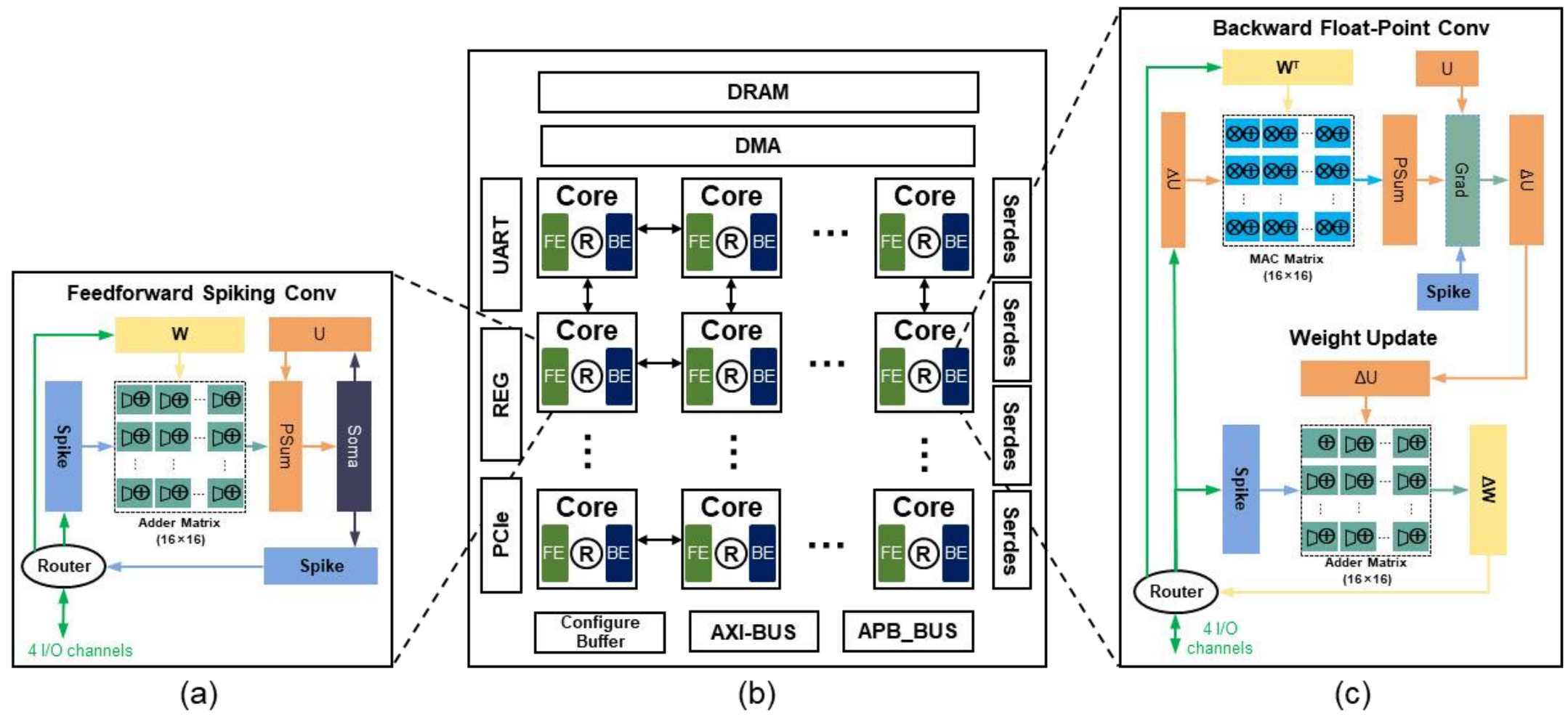

Figure 2. Illustration of a many-core neural network architecture: (a) the structure of the forward core; (b) the many-core chip; (c) the structure of the backward core.

The Feedforward Spiking Convolution in the FP Engine consists of a 16×16 selector and FP16 precision adder. The Backward Float-Point Convolution in the BP Engine is a 16×16 MAC array with FP16 precision. The Weight Update in the WG engine consists of a 16×16 adder array and selectors with FP16 precision. Modules for data transmission in the architecture includes the Dispatch Unit (DU), DMA, and Network On Chip (NOC) consisting of Router and Network Interface (NI). Each computing core contains a Router, and each Router has 6 input and output ports, 2 of which are interconnected with the FP Engine and BP Engine within the core, and 4 are interconnected with 4 adjacent computing cores.

Given the limited storage and computational resources of a single core in the neuromorphic computing system with multiple cores and proximity memory, it is necessary to partition the SNN model prior to model mapping. Following the partitioning, the analysis of data dependencies among the sliced blocks becomes crucial. The training process of the SNN model comprises three stages: forward computation, backward propagation, and weight updating. Both forward computation and weight updating involve binary pulse data as the convolutional operands. As shown in Figure 3, during forward computation, the pulse data $S_t^l$ is convolved with the weights $W^l$. During the model partitioning, we can leverage the characteristic of pulse data transmitting information to reduce interlayer data transmission. It can be observed from Figure 3 that there is frequent data interaction among these three stages. For instance, the pulse data $S_t^l$ generated during forward computation is required for backward propagation, while both the forward $S_t^l$ and the backward $\nabla U_t^l$ are needed for weight updating.

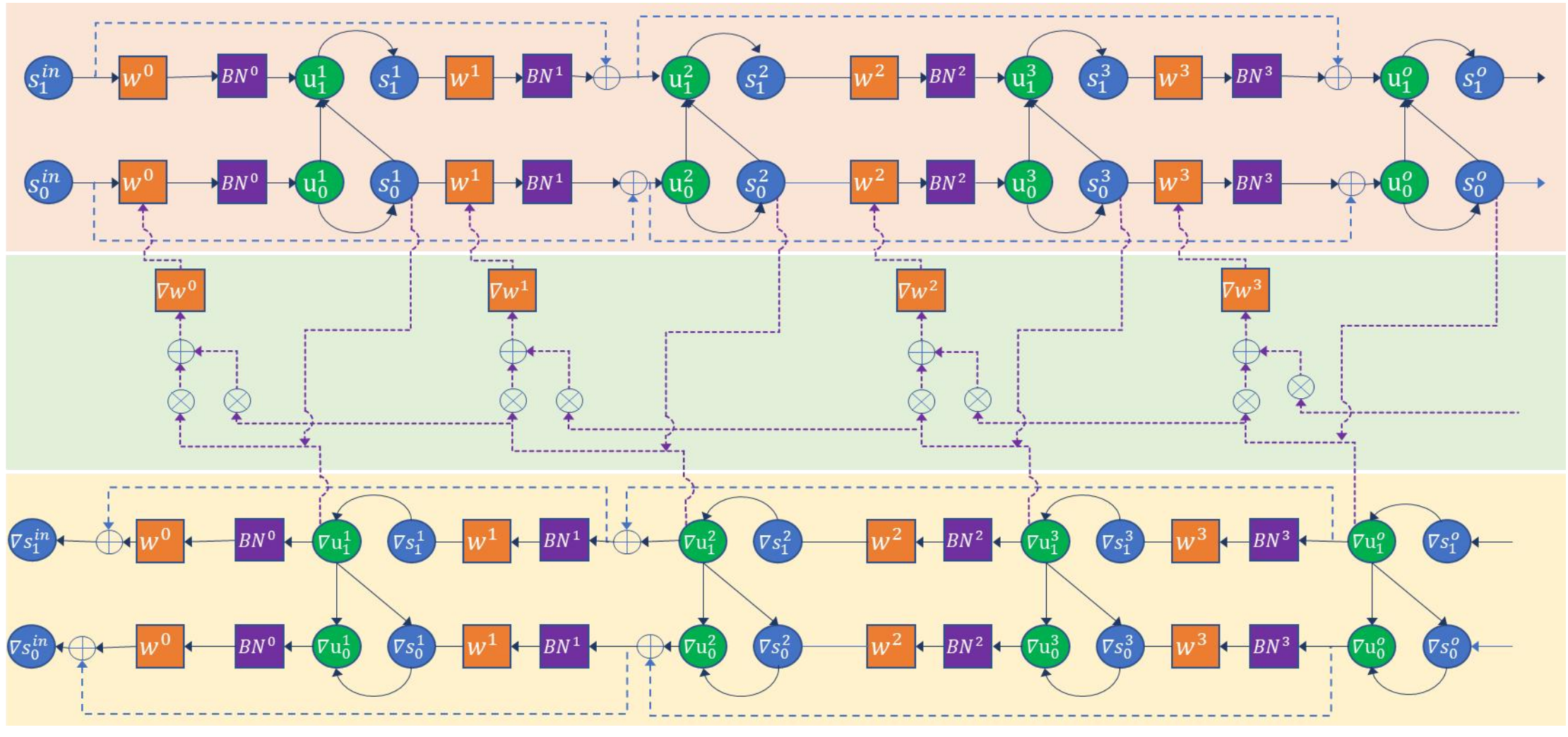

Figure 3. Illustration of the data flow of a neural network.

### 4.2 Model partitioning method

The commonly used neural network partitioning method is mainly based on uniform division of weights. Although Core Placement has further balanced the computational load in the many-core system by evenly dividing the model weights along the input channels $C$ and output channels $K$, the method has limitations when applied to SNN models. As the number of layers in an SNN model increases, the storage space required for its weights and input/output data also increases, while the degree of weight reuse gradually decreases. The uniform weight partitioning method often leads to an uneven distribution of computational tasks, resulting in a short-board effect that significantly impacts the overall efficiency of the many-core system. To address that, we propose a new partitioning method that primarily considers two key factors: the storage requirements of slices and the number of computational operations. First, we conduct a comprehensive evaluation of each layer of the entire model, calculating its computational operations and memory requirements. Then, taking into account the hardware resources of the neuromorphic computing system with multiple cores and proximity memory (such as the on-chip storage size of the computing cores, the number and frequency of computing resources, and the bandwidth of the chip), we perform uneven partitioning along the two dimensions of input channels $C$ and output channels $K$. It approach ensures that the total computation time and data transmission time for each slice are balanced, thus avoiding overburdening or idleness of some computing cores. As shown in Figure 4, we demonstrate our partitioning method using the Spike-ResNet18 model as an example. For the earlier convolutional layers of the model, since the on-chip storage of the computing cores is sufficient to store the allocated slice weights, the partitioning of these slices mainly focuses on computational delay. However, for the later convolutional layers with a larger number of weights, some weights need to be stored in off-chip registers. Therefore, when partitioning these slices, we need to consider the total of both computational delay and weight transmission delay. As can be seen from the Figure 4(d), our balanced partitioning method achieves a significant balance in the distribution of computational and storage latency.

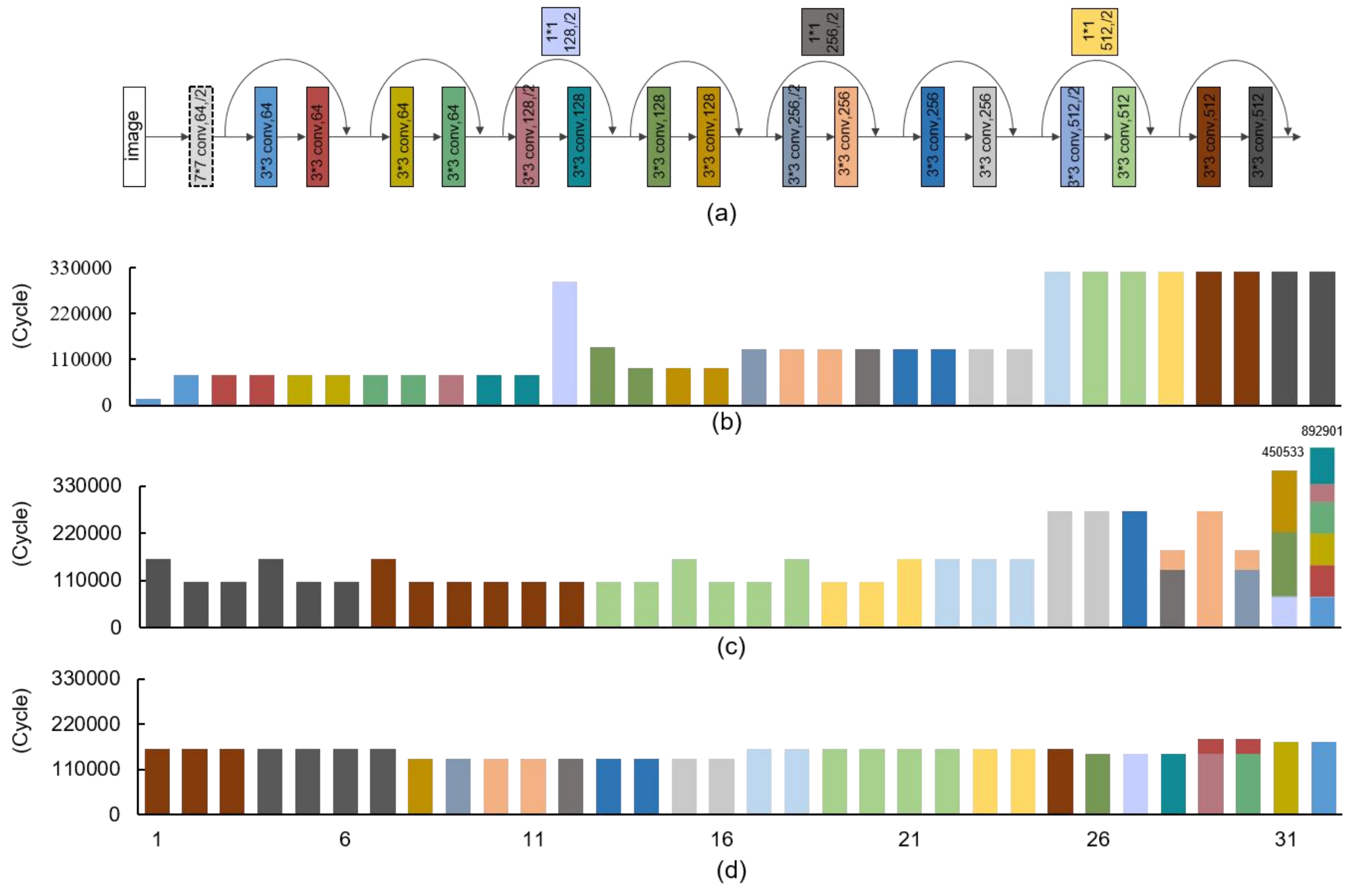

Figure 4. ResNet18 segmentation diagram of three different methods.(a)Network; (b) Computation balance;(c) Storage balance; (d) Computation and storage balance.

### 4.3 Multi -Core Deployment Optimization

The framework of the proposed algorithm is shown in Figure 5. In order to obtain the optimized SNN model, the many-core deployment scheme is trained. We design the Actor Network, model the physical mapping scheme as a coordinate point problem, use graph convolution to fuse the communication connection information of the logic core and the physical architecture, and convert the logic graph into a feature vector. The output of the graph convolutional layer was input into two fully connected layers to learn the complex relationship and combination between different features, the nonlinear mapping ReLU was used as the output layer activation function of the network, and Tanh was used to constrain the output deployment scheme of the network to be less than or equal to the number of computing cores. Actor Network finally outputs an optimal physical mapping scheme. We design a Critic Network to measure the quality of the decision action made by Actor Network, and update the network parameters according to the error, so that the whole ensemble algorithm has a relatively good performance. In addition, the actions output by our designed algorithm will be input into the Actor Network with updated weights again, which reduces the number of iterations of the network to some extent.

We have developed an algorithm framework aimed at optimizing the deployment scheme of SNN models in a multicore system. we introduced an Actor Network to model the physical mapping scheme as a coordinate point problem. To fully integrate the communication connection information and physical architecture of the logical cores, we have employed graph convolution techniques to convert the logical graph into feature vectors. These feature vectors are then fed into two fully connected layers to learn the complex relationships and combinations among different features. We chose the nonlinear mapping ReLU as the activation function and used the Tanh function to ensure that the output deployment scheme does not exceed the limit of the number of compute cores. Eventually, the Actor Network will output an optimal physical mapping scheme. To evaluate the decision-making effectiveness of the Actor Network, we have also designed a Critic Network. The network is capable of assessing the quality of decisions made by the Actor Network and updating the network parameters based on errors to ensure the overall performance of the integrated algorithm. Furthermore, in our algorithm design, the output actions are fed back into the Actor Network as inputs for weight updates which reduces the number of iterations required for the network and improves the efficiency of the algorithm.

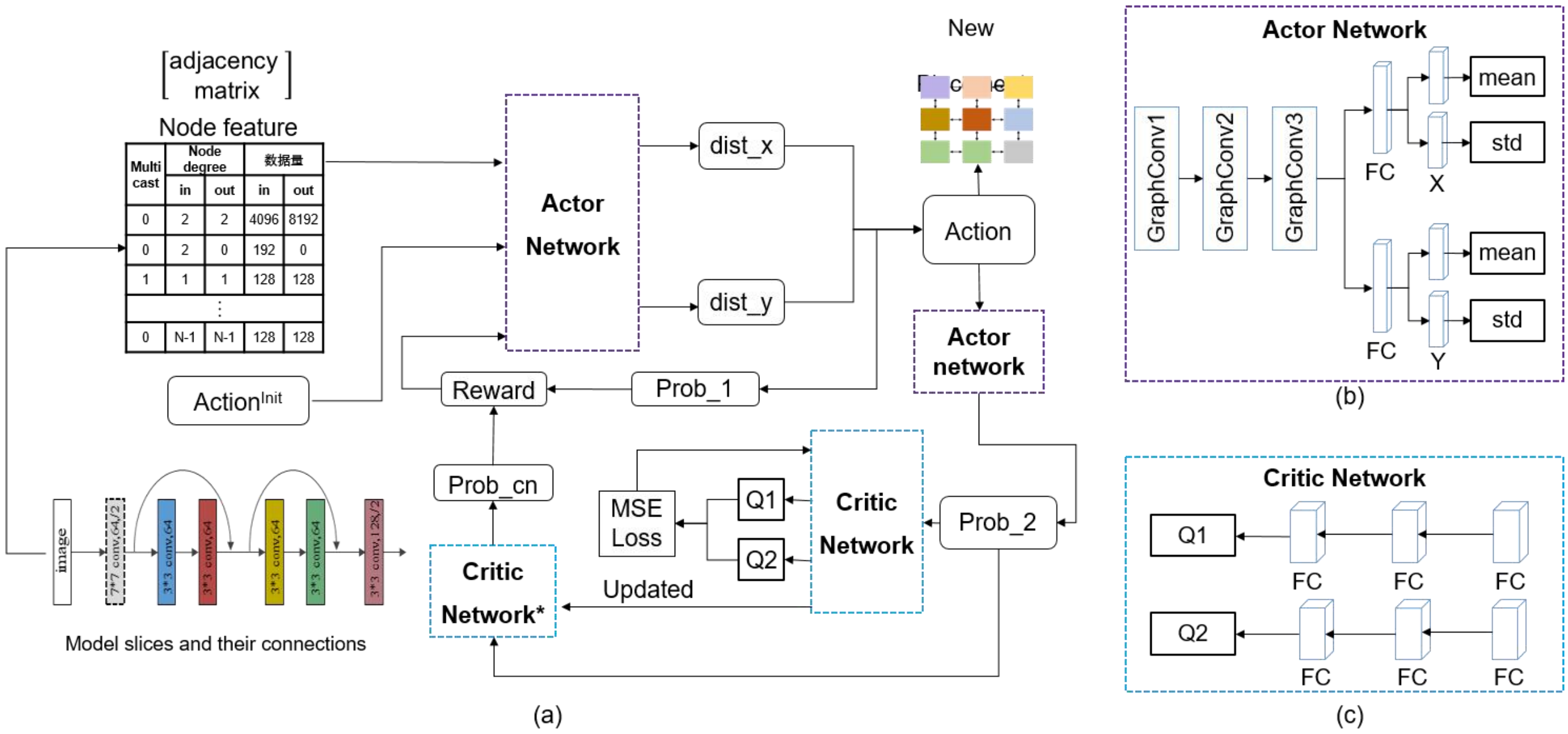

Figure 5. Overview of our approach: (a) a detailed overall schematic of our approach; (b) structure of the Actor Network; (c) structure of the Critic Network.

**State environment.** During the modeling process, the network graph of the logical cores and the feature representations of each logical node are first transformed into an information format that the Actor Network can process. We represent the logical graph as an adjacency matrix and further normalize it into a Laplacian matrix. Five variables are used to represent the attributes of the node features in five different dimensions. As a simple example shown in Figure 5, the Node feature represents the characteristics of the nodes, the Multi-cast is a boolean value indicating whether the node is a multicast node, the node degree is divided into two directional features: $in$ represents the flow of other nodes to this node, and $out$ represents the outflow direction of this node. The data volume is also divided into two sub-features, where the value $in$ in represents the amount of data received by the node, and $out$ represents the amount of data sent by the node. After the graph convolution operation, the logical graph and the node attributes are encoded into vector data, which serves as the state environment. Notably, this environmental representation remains unaltered throughout the model training process.

**Action.** The output of the Actor Network represents the mapping of nodes in the logical graph to positions in the many-core system. After the graph convolution operation, we utilize two fully connected layers to conduct a comprehensive search for deployment solutions in both the horizontal and vertical dimensions. For example, in the mapping of Spike-ResNet18, we divide the model into 32 logical nodes and map them to a $4 \times 8$ multicore system. The output of the Actor Network is four $32 \times 32$ vectors, representing the mean and variance of each logical node in the two-dimensional array direction. We apply Gaussian distributions to the outputs of both dimensions and clip them to the range of [-*x*, *x*]. Then, each dimension is discretized equidistantly into $4 \times 8$ discrete values. The combination of discrete values at the same position in both dimensions represents the probability of mapping a logical node to different physical cores. When quantifying probabilities into deployment solutions, if multiple logical nodes are mapped to the same physical core, we search for idle physical cores by rotating on the axis with the minimum step distance in a clockwise direction, the logical nodes are mapped in order of their priority to the first position with the smallest Manhattan distance from the orange core during the clockwise search.

**Reward.** In order to search the optimal deployment scheme for many-core training of SNN model, reduce inter-core data communication transmission, eliminate on-chip network communication hotspots, minimize model training time and power consumption, and improve computing resource utilization. In the design of reinforcement learning model, we will consider the above factors at the same time. The power consumption and delay are linear with respect to communication cost, so the reward function of Actor Network only needs to consider the communication between computing cores. According to Policy, we design the reward function as follows:

$$\mathcal{J}_\theta = \max\{-\left(CDV_{left} + CDV_{right} + CDV_{up} + CDV_{down}\right)\} \tag{4}$$

Where $CDV(left, right, up, down)$ represents the amount of communication data between the computing core and the computing cores in the up, down, left, and right directions.

**Weight Update.** we use the power of indirectly supervised deep reinforcement learning as a direct nonlinear optimizer, while also optimizing the surrogate loss with PPO and the clipping objective to update the parameters of the Actor Network, as follows:

$$\mathcal{L}^{CLIP}(\theta) = \widehat{E_t}\left[\min\left(r_t(\theta)\widehat{A_t}, clip(r_t(\theta), 1-\epsilon, 1+\epsilon)\widehat{A_t}\right)\right] \tag{5}$$

The graph convolutional layer in the Actor Network is a pre-trained Network, which does not need to be updated in the optimization, and the mean square error is used to optimize the Critic Network.

## 5 EXPERIMENT

### 5.1 Experimental Details

The experiment data set is segmented by different numbers of logical cores of Spike-ResNet18, Spike-vgg16 and Spike-ResNet50 models. In this paper, 32 cores and 64 cores are segmented respectively to accurately obtain the connection relationship graph and data traffic between nodes. A total of six datasets were formed to evaluate our proposed deployment optimization algorithm. For the structure and hyperparameters in the proposed method, the feature size in the graph convolutional layer was set to 32, the clipping-range and batch size were set to 0.1 and 256, and the learning rate was set to 0.005. ppo_epoch and and ppo_clip are set to 10 and 0.5, respectively, and the reward value is trimmed to $[-10,10]$. All the results are obtained by simulator evaluation based on the many-core near-memory brain-like computing architecture described in this paper.

Compare the proposed method with the following deployment methods:

(1) Sequential deployment: Sequential deployment includes two deployment modes. One is to deploy logical nodes one by one from the top left corner, which is represented by Zigzag, and this method is used as the comparison baseline in the subsequent comparison. The other is to deploy the logical nodes from the first physical core to the nearest row, which is represented by Sigmate.

(2) Random search: This algorithm is an optimization method with large search range but low efficiency. In the experimental search, the performance evaluation index and function are the same as the proposed method based on DDPG. One deployment scheme is randomly searched for evaluation each time. The scheme with the minimum communication cost consumption is selected as the output, and the random search method is denoted by RS.

### 5.2 Deployment Optimization for Many-core Computing Systems

Based on the method proposed in this paper, the deployment scheme of each model is simulated on the above experimental platform, and the communication cost, delay and communication local hot spot are compared with the current basic deployment method. Figure 6 shows the experimental data corresponding to the optimal deployment schemes of different models for each method in the case of 32 cores, including the cost of communication, latency, throughput and hop distance distribution, where the Batch Size is 8 in model inference and training. Figure 6(a) shows the communication cost of Spike-ResNet18, Spike-vgg16 and Spike-ResNet50. During model inference, the proposed method is reduced by 34.12%, 50.68%, and 32.77% respectively compared with the three basic methods. In training, the communication cost of the proposed method is reduced by 28.84%, 32.92% and 18.89%, respectively. Unlike inference, the optimization of communication cost is more complex and less optimized due to the high dependence of model data in training, and it can also be seen that the model with more complex structure optimizes better. Figure 6(b)-(c) show the latency and throughput of the chip. In inference, the latency and throughput of the proposed method is reduced by about 8%, and during training it is reduced by about 10%. Figure 6(d) shows the hop distribution of all connections after mapping logical nodes of each model. In the case of the same number of transmission packets, the communication cost is lower with fewer transmission hops. From the hop count distribution, it can be seen that The hop distance of the proposed method deployment scheme is concentrated in a small range, while the other methods are more discrete. Compared with the zigzag deployment scheme, the average hop count of each data packet of the proposed method is reduced by 0.67, respectively. These results show that the interconnected logical nodes are mapped in adjacent positions. All packets are transmitted in a minimal local area, implicitly reducing the overall system runtime and power consumption. Figure 7(a)-(b) shows the 32-core communication hotspot map of the basic method and the proposed method, where the darker red indicates the larger communication load of this physical core. The figure shows the distribution of communication hotspots corresponding to different deployment methods of the three models. Whether it is model training or inference, the communication data volume distribution of the proposed method is significantly better than the other three methods, and there is basically no physical core with large communication volume in the whole chip system.

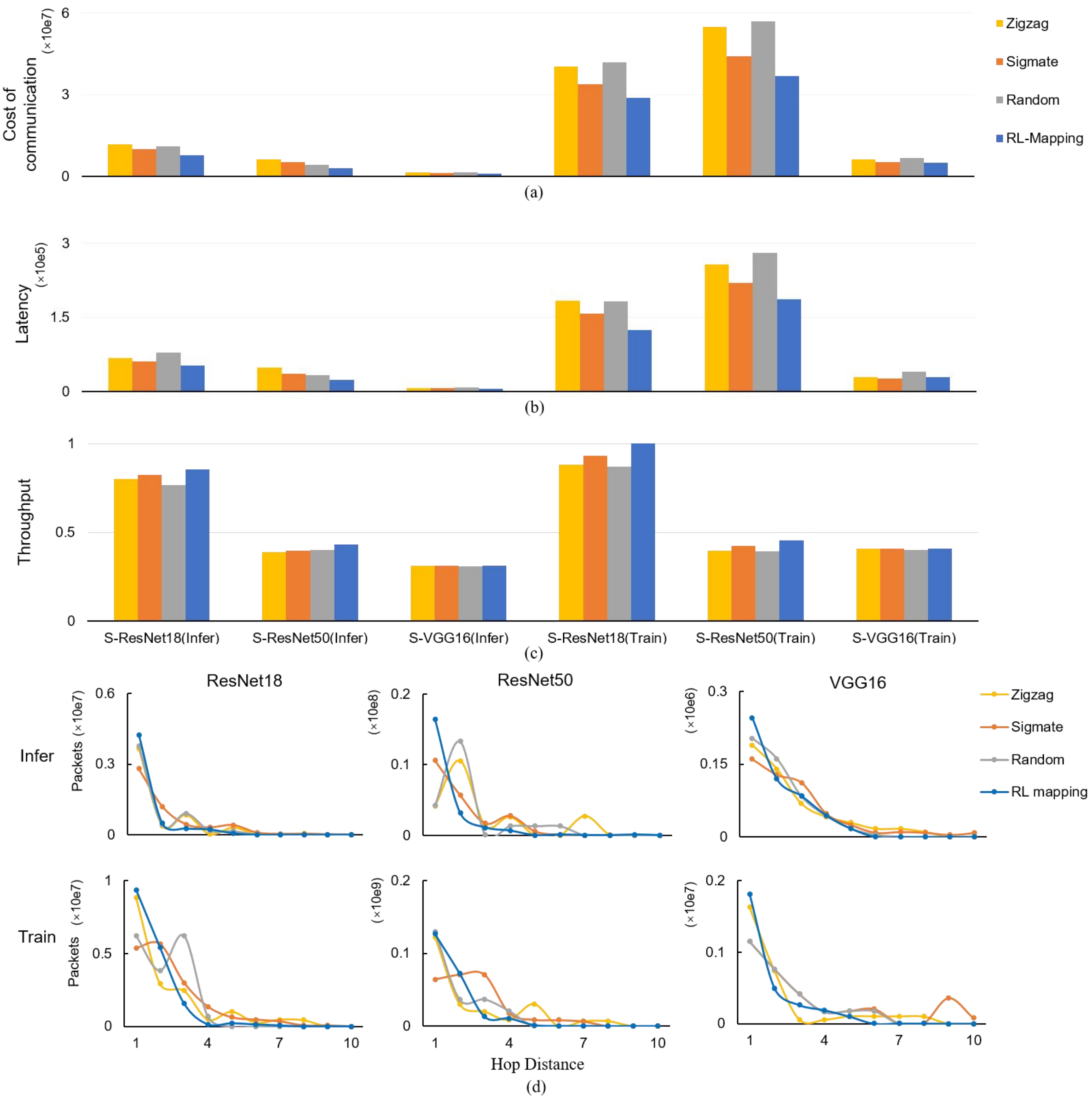

Figure 6. Experimental data of a 32-core chip system: (a) distributions of cost of communication; (b) distributions of latency; (c) distributions of throughput; (d) distributions of hop distances of core placements placed.

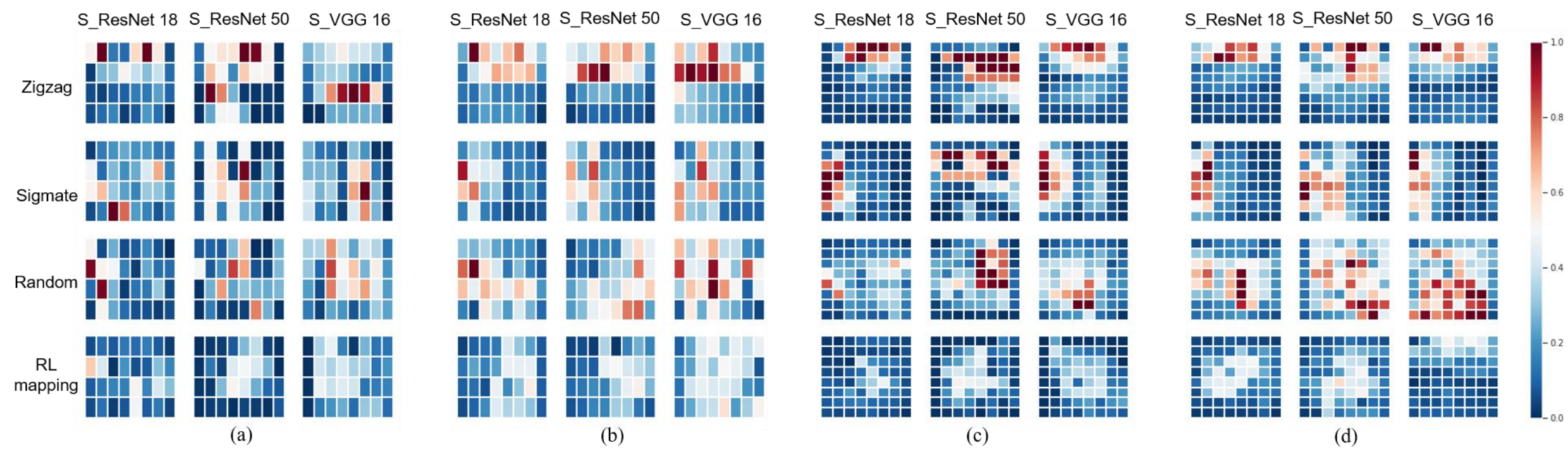

Figure 7. Visualization of the number of packets transmitted on each core in a time phase, the color of each core represents the packets are transmitted by each node, packets of each experiment are normalized. (a) 32 core inference; (b) 32 core training; (c) 64 core inference; (d) 64 core training.

In order to further explore the generalization ability of the method proposed in this paper, a 64-core many-core simulation platform is designed and built. The Spike-ResNet18, Spike-vgg16 and Spike-ResNet50 models are divided into 64 logic cores by using the method which we proposed. Three base deployment methods and reinforcement learning-based methods were used for mapping optimization, respectively. Figure 8 shows the results of the experiment. It shows that our method is also applicable to different many-core systems. In inference and training, the communication cost of the proposed method is reduced by more than 22.64%, the amount of data communication between cores is also significantly balanced, the latency is reduced by more than 13%, S-ResNet18-50 is even more is 30%, and the throughput of S-ResNet18-50 are increased by about 13%, though the S-VGG16 improvement is not obvious.

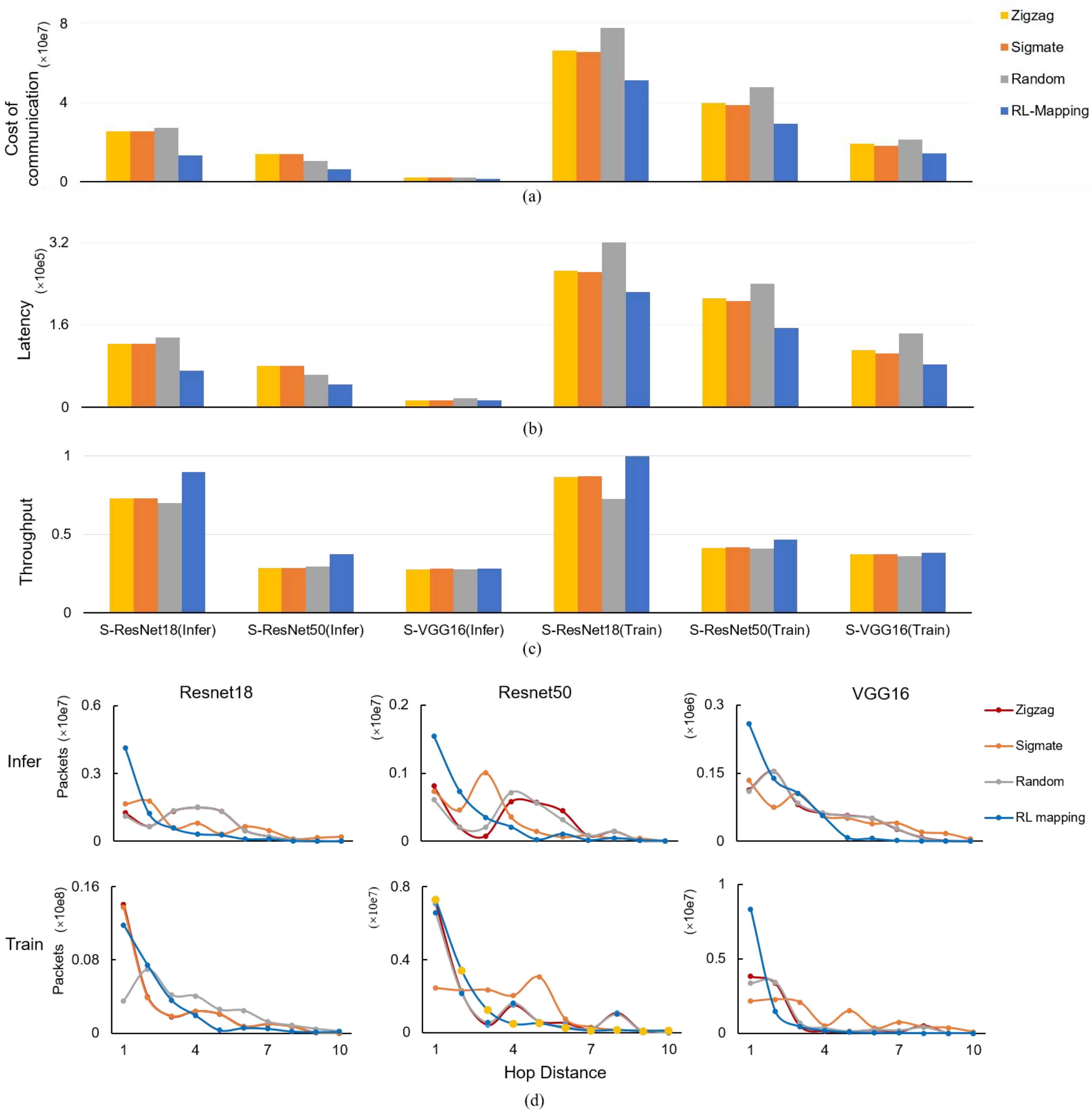

Figure 8. Experimental data of a 64-core chip system. (a) distributions of cost of communication; (b) distributions of latency; (c) distributions of throughput; (d) distributions of hop distances of core placements placed.

We also studied the computational core utilization of the proposed method on different computation pipeline algorithms between cores. Figure 9 shows the waveform of computational core utilization variation in different clock cycles of the common layer-wise and the computation pipeline method in FPDeep. The layer-wise does not proceed to the next computation core until the tasks on the previous computation core are completely completed, and the FPDeep means that when the input of the next computation core is calculated by the previous computation core, the two computation cores will be in the computation state at the same time. Figure 9(a) shows the computing resource usage in layer-wise, and it is easy to find that most of the forward and reverse computing cores are completely idle in the clock cycle shown; Figure 9(b) shows the resource utilization of FPDeep. Since it can be started without waiting for the completion of the previous computing core task, the clock cycle of FPDeep to complete one round of training is significantly reduced, and the number of cores in computing state in the same clock cycle is significantly increased. Figure 9(c) shows the resource utilization curves of the two methods, there is a significant increase in computational resource utilization of FPDeep.

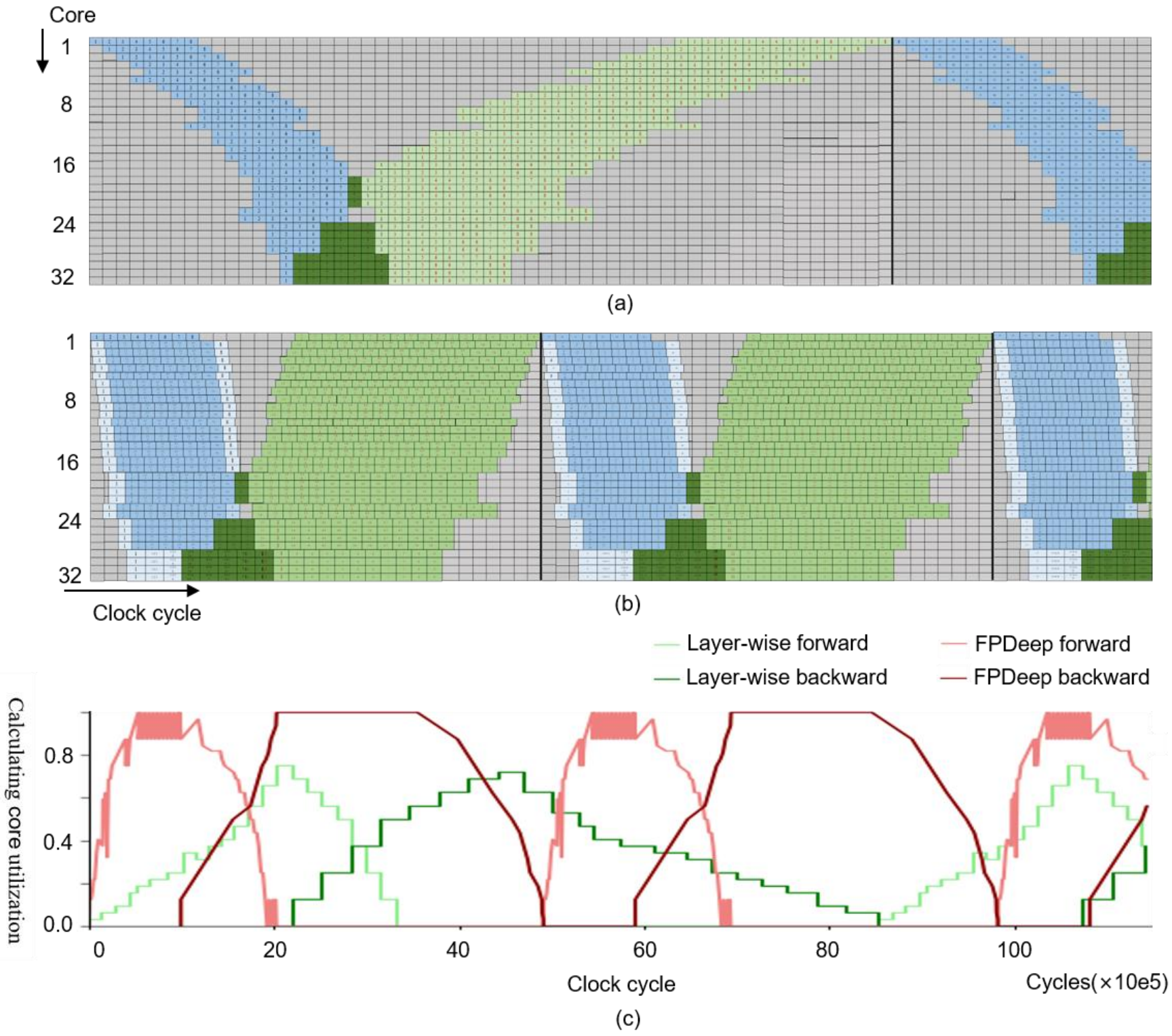

Figure 9. The computing resource usage Layer-wise and FPDeep. (a) Layer-wise ; (b) FPDeep; (c) Computational Resource Utilization.

### 5.3 Comparison with existing methods

Tianjic is a many-core artificial intelligence chip, which consists of a 2D mesh on-chip network-on-chip (NoC), an on-chip router for off-chip communication, and necessary chip peripherals. Policy proposes an optimized deployment method based on reinforcement learning for the inference mapping optimization of ANN models. The proposed method successfully reduces model inference running time, communication cost, and average traffic load between cores. We build an experimental simulation platform based on the known data of Tianjic movement to compare the effects of the deployment optimization method proposed in this paper and the Policy method. This section mainly arranges two sets of experiments for verification analysis, one is the model inference deployment optimization based on the Tianjic, and the other is the model training deployment optimization based on the many-core architecture of this paper.

In the experiment, the logic maps of ResNet18, VGG16 and ResNet50 are used as data sets to verify the practicability of the proposed method for ANN models and the optimization effect of Policy. Experiments are carried out on a single-chip system with 32 and 64 cores, and Figure 10 shows the experimental effects of Zigzag deployment, Policy reinforcement learning deployment, and the proposed method. Figure 10(a)-(b) show the distribution of communication cost for model inference and training respectively. In inference, the proposed method has a significant reduction compared with Zigzag and Policy, and the average communication cost is reduced by 43.5% and 6.5%, respectively. The average communication cost during model training is reduced by 29.2% and 8.7%, respectively. Figure 10(c) shows the distribution of data transmission hops of all logical node connections after physical mapping. In reasoning, the proposed method has better optimization effect on data packet transmission distance than the other two methods, and the average transmission distance of data packet is reduced by 1.128 and 0.167 respectively, which is 0.167 less than the average transmission distance of single data packet of the Policy method. During training, the packet transmission distance is reduced by 0.636 and 0.240 on average, respectively. Figure 11 shows the inter-core data communication heat maps of different deployment methods, and the deeper the red color is, the greater the communication volume. It can be seen that our method is better than the Policy method in balancing inter-core communication.

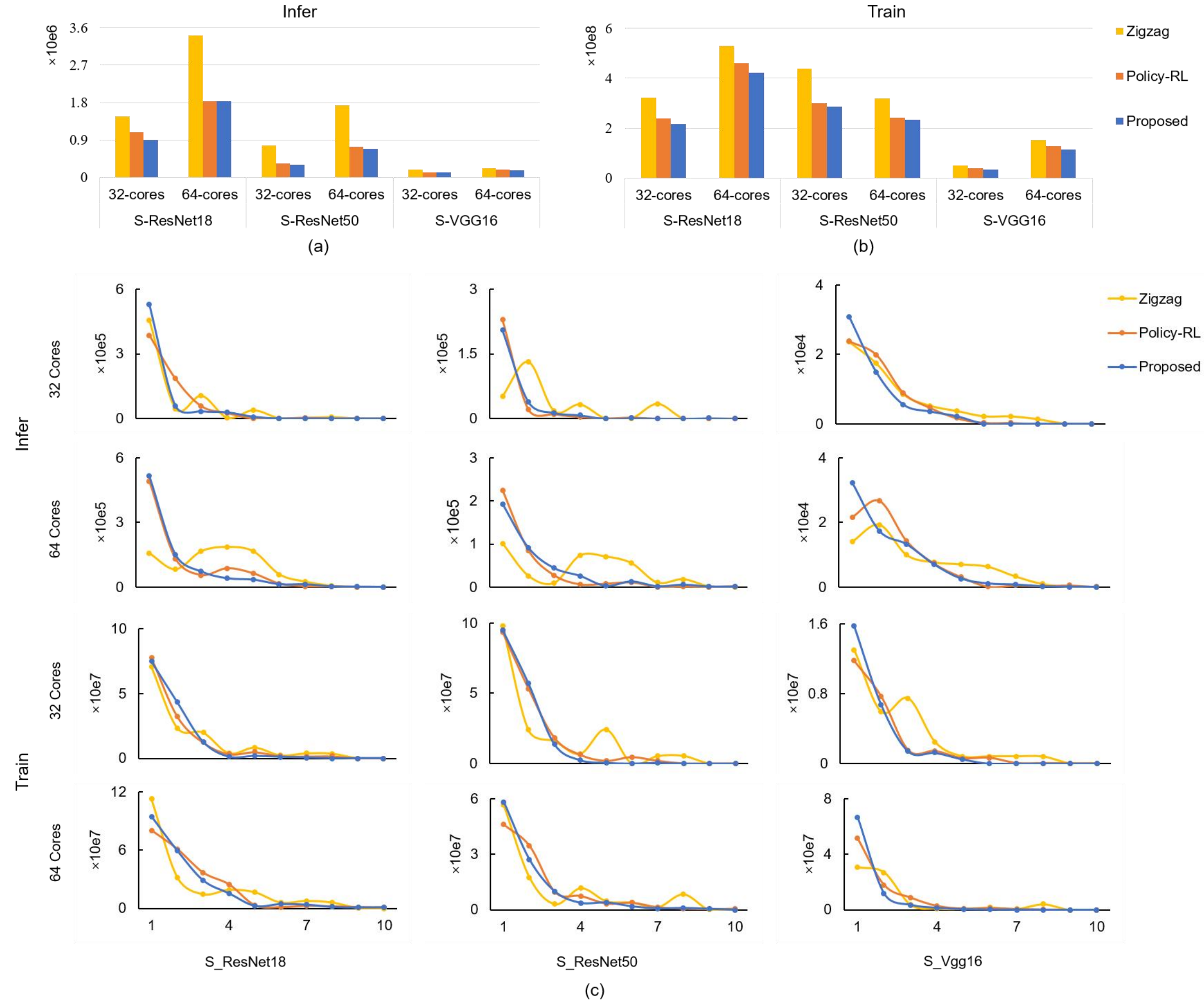

Figure.10 Experimental data of two SP methods and our agent: (a) cost of communication for inference; (b) cost of communication for training (c) distributions of hop distances of core placements placed.

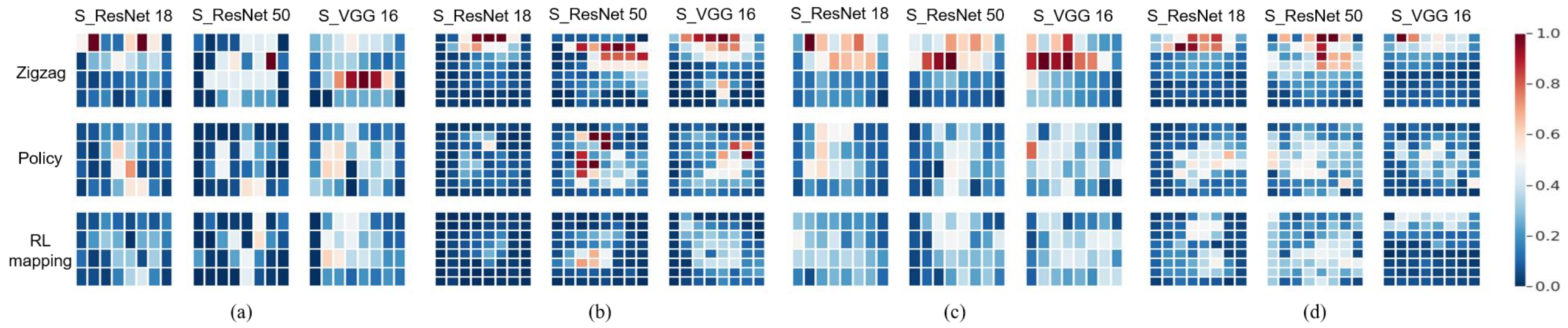

Figure 11. Visualization of the number of packets transmitted on each core in a single time phase for Zigzag, policy and RL-mapping algorithms. (a) 32 core inference; (b) 64 core inference; (c) 32 core training; (d) 64 core training.

Experiments show that the proposed method can reduce the power consumption of the system by optimizing the mapping scheme, balancing the inter-core communication hotspots and reducing the number of single packet transmission hops, and the deployment optimization effect of ANN model inference and SNN model training is better than the existing methods.

## 6 CONCLUSION

In this paper, we focus on the deployment optimization of spiking neural network training and propose a many-core deployment optimization method for SNN network training based on reinforcement learning. Current research mainly focuses on the inference deployment optimization of ANN chip architecture, but lacks in-depth research on model training. Considering the problems of local communication hotspots between cores, low utilization of computing cores and low system throughput in many-core near-memory brain-like computing systems during model training and mapping, we first propose a model segmentation method for balancing storage and computation strategies. Secondly, a many-core deployment optimization method for training spike neural networks is proposed based on Off-policy Deterministic Actor-Critic. The Actor network and Critic network are designed, and the continuous values matching the number of cores are used as the output of the policy network. Then, they are discretized again to obtain a new deployment scheme, and the parameters of the policy network are updated through the proximal policy optimization to realize the deployment optimization of SNN model in many-core near-memory brain-like computing architecture model training to reduce chip power consumption. We use Spike-ResNet18, Spike-VGG16 and Spike-ResNet50 models to validate and evaluate the proposed method, and the deployment effect of the proposed method is better than that of Zigzag, Sigmate, RS and Policy deployment methods.